\documentclass[twocolumn]{aastex631}

\usepackage{subfigure}

\begin{document}

\title{No signature of the birth environment of exoplanets from their host stars' Mahalanobis phase space}

\author[0000-0002-5710-7445]{George A. Blaylock-Squibbs}\thanks{Email: gablaylock-squibbs1@sheffield.ac.uk}
\affiliation{Department of Physics and Astronomy, The University of Sheffield, Hounsfield Road, Sheffield, S3 7RH}

\author[0000-0002-1474-7848]{Richard J. Parker}\thanks{Royal Society Dorothy Hodgkin fellow}
\affiliation{Department of Physics and Astronomy, The University of Sheffield, Hounsfield Road, Sheffield, S3 7RH}


\author[0000-0002-3767-8187]{Emma C. Daffern-Powell}
\affiliation{Department of Physics and Astronomy, The University of Sheffield, Hounsfield Road, Sheffield, S3 7RH}



\begin{abstract}
The architectures of extrasolar planetary systems often deviate considerably from the `standard' model for planet formation, which is largely based on our own Solar System. In particular, gas giants on close orbits are not predicted by planet formation theory and so some process(es) are thought to move the planets closer to their host stars. Recent research has suggested that Hot Jupiter host stars display a different phase space compared to stars that do not host Hot Jupiters. This has been attributed to these stars forming in star-forming regions of high stellar density, where dynamical interactions with passing stars have perturbed the planets.  We test this hypothesis by quantifying the phase space of planet-hosting stars in dynamical $N$-body simulations of star-forming regions. We find that stars that retain their planets have a higher phase space than non-hosts, regardless of their initial physical density. This is because an imprint of the kinematic substructure from the regions birth is retained, as these stars have experienced fewer and less disruptive encounters than stars whose planets have been liberated and become free-floating. However, host stars whose planets remain bound but have had their orbits significantly altered by dynamical encounters are also primarily found in high phase space regimes. We therefore corroborate other research in this area which has suggested the high phase space of Hot Jupiter host stars is not caused by dynamical encounters or stellar clustering, but rather reflects an age bias in that these stars are (kinematically) younger than other exoplanet host stars. 

\end{abstract}

\keywords{star forming regions (1565) -- exoplanet migration (2205) -- stellar dynamics (1596) -- $N$-body simulations (1083)}

\section{Introduction} \label{sec:intro}

Observations and theory have shown that stars form in groups with tens, to tens of thousands, of stellar siblings, with a wide range of stellar densities that exceed those in the Galactic disc by several orders of magnitude \citep{lada_embedded_2003,bressert_spatial_2010}. 

Star-forming regions exhibit spatial and kinematic substructure \citep{gomez_spatial_1993,larson_star_1995,cartwright_statistical_2004,sanchez_spatial_2009,andre_filamentary_2014}, which is erased over time though dynamical evolution \citep{klessen_mean_2001,parker_characterizing_2012}, sometimes solely two-body relaxation, or a combination of this and violent relaxation.

Furthermore, observations show that planet formation occurs contemporaneously with star-formation \citep{2001ApJ...553L.153H,alves_star_planet_formation_simultaneous_2020}, such that the star-formation environment may strongly influence the planet formation process, depending on the density of the star-forming region and how long the planets' host stars spend in the region \citep{parker_birth_2020}.



In the lowest density star-forming regions ($\tilde{\rho} \sim 10$s\,M$_\odot$\,pc$^{-3}$), if massive stars are present (e.g.\,\,in OB associations) then photoionising radiation will lead to the evaporation of gas (and to a lesser extent, dust) from the discs \citep{scally_destruction_2001,2018MNRAS.475.5460H,2018MNRAS.481..452H}, possibly hindering the formation of gas giant planets \citep{clarke_role_1991,nicholson_rapid_2019,concha-ramirez_external_2019} and altering the structure of the dust in the discs \citep{2020MNRAS.492.1279S}. 

In moderately dense star-forming regions ($\tilde{\rho} \sim 100$s\,M$_\odot$\,pc$^{-3}$, which most nearby star-forming regions are at least as dense, if not more so at birth, \citealp{parker_dynamical_2017,parker2022}) planetary orbits can be disrupted. This can be either a relatively minor change in their semi-major axis and/or eccentricity, or a more energetic interaction that leads to the planets becoming unbound from their original host stars and becoming a free-floating within the star-forming region \citep{kobayashi_and_ida_effects_of_stellar_encounter_on_disc_2001, daffern-powell_great_2022}. 

In the most dense star-forming regions ($\tilde{\rho} > 1000$\,M$_\odot$\,pc$^{-3}$) protoplanetary discs can even be truncated by interactions with passing stars \citep{scally_destruction_2001,2006ApJ...642.1140O,2012A&A...538A..10S,2016ApJ...828...48V,2018MNRAS.478.2700W}, potentially limiting the material available for planet formation in the outer discs.

A significant unanswered question in astrophysics is to what degree does the star-forming environment shape the architectures of planetary systems? For instance, are Hot Jupiters \citep{dawson_and_johnson_hot_jupiter_origins_2018} formed from the migration of planets in discs \citep{2003ApJ...588..494M}, or are they produced later via dynamical instabilities in the planetary systems caused by interactions with other stars \citep{2003ApJ...589..605W,2024MNRAS.527..386L,2024arXiv240111613B}?

However, most exoplanet host stars (and indeed our own Sun) are main sequence stars in the Galactic field, and are likely to have migrated from their birth star-forming regions. There appears to be little prospect of identifying the dynamical signatures of planet host stars' birth environments. 




However, \citet{winter_stellar_2020} used \emph{Gaia} observations to build up a 6D phase space (i.e.\,\,position and velocity) distribution for exoplanet host stars, and quantified the phase space around stars using the dimensionless \citet{mahalanobis_generalized_1936-2} distance. \citet{winter_stellar_2020} then use the Mahalanobis distance to define a phase space `density'\footnote{This phase space `density' is not the physical volume or surface density of the stars, but the two may be related as postulated by \citet{winter_stellar_2020}.}. 

After finding the relative Mahalanobis `densities' of exoplanet host stars in their data, \citet{winter_stellar_2020} categorize each exoplanet host star according to its local Mahalanobis phase space; either low, ambiguous or high. In a low Mahalanobis phase space nearby stars will have very different velocities and \citet{winter_stellar_2020} call these `field stars'; though technically all of the stars in their samples are Galactic field stars. In a high Mahalanobis phase space stars will exhibit clustering in their position--velocity phase space, and \citet{winter_stellar_2020} suggest that the high Mahalanobis phase space  regime indicates the imprint of a dense, clustered birth environment. 

\citet{winter_stellar_2020} find that Hot Jupiter host stars are predominantly found in high Mahalanobis phase space, which they interpret as being due to them forming in high density star-forming regions, where interactions with passing stars led to the formation of Hot Jupiters.  
However, several authors \citep{adibekyan_stellar_2021, mustill_hot_2022} have contended that the high Mahalanobis phase space is due to an age bias. Hot Jupiters tend not to survive due to orbital decay and photoevaporation of their atmospheres \citep{hamer_et_al_hot_jupiter_hosts_are_younger_2019,chen_et_al_hot_jupiter_host_stars_younger_2023}. Younger stars tend to be kinematically `cooler', which results in the stars being closer to each other in the 6D Mahalanobis phase space. 


In this work we investigate whether stars whose planetary systems  are perturbed in their birth star-forming regions retain an imprint of the initial stellar clustering. We follow the evolution of the 6D Mahalanobis phase space for sets of $N$-body simulations where the stars host Jupiter mass planets. We then look for correlations between stars in high Mahalanobis phase space and the orbital architectures of their planetary systems. The paper is organised as follows. We outline our methods in Section~\ref{methods}, we present our results in Section~\ref{results} and we provide a discussion in Section~\ref{discuss}. We conclude in Section~\ref{conclude}. 

\section{Methods}
\label{methods}
In this section we summarise the setup of the $N$-body simulations and then discuss how we quantify the 6D Mahalanobis phase space metric.

\subsection{Simulation set up}
We use four sets of 20 subvirial (virial ratio $\alpha_{\rm{vir}} = 0.3$, where $\alpha_{\rm{vir}} = 0.5$ is virial equilibrium) simulations from \citet{daffern-powell_great_2022} containing 1000 stars drawn from a \citet{maschberger_function_2013} IMF. 500 of these stars are randomly assigned a single Jupiter-mass planet at either 5\,au or 30\,au. The orbital eccentricities and inclinations are all set to zero. The average mass of host stars in the simulations is $\sim 0.5\,$M$_\odot$, which is lower than the observed mean Hot-Jupiter host star mass of $1\,$M$_\odot$ \citep{schneider_exoplanetseu_2011}. 
Their is a small dependence of planet perturbation on stellar mass in these simulations \citep[][who find that planets are more likely to be disrupted around M-dwarfs than G-dwarfs]{Parker_binary_and_planet_system_destruction_2023}. However, this subtle difference (of order 10\%) would be difficult to observe in reality.

To test if different densities of the simulations results affects the numbers of planet-hosting stars in high Mahalanobis phase space regimes we use low and high density simulations with local median volume densities of $\sim 10^{2}\,$M$_{\odot}\,$pc$^{-3}$ and $\sim 10^{4}\,$M$_{\odot}\,$pc$^{-3}$, respectively. 

For additional information on the simulations see \citet{daffern-powell_great_2022}. We summarise the simulations in Table~\ref{table:simulation initial conditions}.

\begin{table}
	\centering
	\caption{Table summarising the simulations used in this work. From left to right the columns are the fractal dimension $D_f$, the median local stellar density of the simulations $\tilde{\rho}$, the virial ratio $\alpha_{\rm{vir}}$, the number of stars $N_{\star}$ and the initial semi-major axis of the Jupiter mass planets' initial orbits, $a_{\rm{i}}$.}
	\label{table:simulation initial conditions}
	\begin{tabular}{lcccc} %
		\hline
		$D_{f}$ & $\tilde{\rho}$ (M$_{\odot}\,$pc$^{-3}$) & $\alpha_{\rm{vir}}$ & $N_{\star}$ & $a_{\rm{i}}$ (au)\\
		\hline
		1.6 & $10^{2}$  &  0.3   & 1000 & 5  \\
		1.6 & $10^{2}$  &  0.3   & 1000 & 30  \\
        1.6 & $10^{4}$  &  0.3   & 1000 & 5 \\
        1.6 & $10^{4}$  &  0.3   & 1000 & 30 \\

		\hline
	\end{tabular}
\end{table}

\subsubsection{Dynamical evolution}
The simulations are evolved using the \texttt{kira} integrator, a part of the \texttt{starlab} software package \citep{zwart_star_1999, portegies_zwart_star_2001}. The \texttt{kira} integrator uses a 4$^{\rm{th}}$-order Hermite scheme with individual time steps. We limit our planetary systems to one Jupiter-mass planet and therefore treat the star--planet systems as a binary \citep{zwart_star_1999, portegies_zwart_star_2001}.

\subsubsection{Dynamical timescales}
The simulations are run for 10\,Myr which allows for sufficient dynamical evolution and relaxation to occur. 
The two stellar density regimes in the simulations result in different crossing times, which in turn results in different relaxation times. The high-density ($\tilde{\rho} = 10^4$\,M$_\odot$\,pc$^{-3}$) regions have shorter crossing times of around 0.1\,Myr, whereas the low density ($\tilde{\rho} = 10^2$\,M$_\odot$\,pc$^{-3}$) regions have crossing times on the order of 2\,Myr. Since both the crossing times for the simulations are well below 10\,Myr we are able to simulate the dynamical interactions between stars and planets to suitable degree. The longer crossing time for the low density regions results in the spatial and kinematic substructure being present for longer, although it is still erased within 5 Myr. 

Our choice of 10\,Myr for the end-time  of simulations is also informed by observations that highlight the paucity of star-forming regions remaining bound beyond 10\,Myr \citep{lada_embedded_2003, kruijssen_fraction_2012}.



\subsection{Mahalanobis phase space metric}
The Mahalanobis phase space `density' metric uses the Mahalanobis distance introduced by \citet{mahalanobis_generalized_1936-2} to measure the distance of a point from a distribution. The Mahalanobis distance allows for the comparison of high dimensional data sets (e.g. position-velocity phase space is 6D) in which each parameter can be very different in physical scale from one another (i.e. positions measured in pc and velocities in km\,s$^{-1}$). The Mahalanobis distance between two points (in this case stars) is
\begin{equation}
     m_{\rm d}(\vec{x}, \vec{y}) = \sqrt{\left(\vec{x}-\vec{y}\right)^{\rm{T}} \textbf{S}^{-1} \left(\vec{x}-\vec{y}\right)},
\end{equation}
where $\vec{x}$ and $\vec{y}$ are vectors containing the position and velocity information for two stars and $\textbf{S}^{-1}$ is the inverse of the covariance matrix. By multiplying the data set by $\textbf{S}^{-1}$ it is variance normalised and the units are removed. This rescaling means that moving one unit along on one axis in 6D Mahalanobis space is the same as moving one unit along any other axis in the 6D Mahalanobis space.

\citet{winter_stellar_2020} define the Mahalanobis phase space `density' (we hereonin refer to this as a `metric' rather than a `density' to avoid confusion with the volume density of the stars) as  
\begin{equation}
     \rho_{\rm{m,N}} = N \, m_{\rm d, N}^{-D_{\rm p}},
     \label{mahalanobis_eqn}
\end{equation}
where $N$ is the nearest neighbour number, $m_{\rm d, N}$ is the Mahalanobis distance to the $N^{th}$ nearest neighbour and $D_{\rm p}$ is the number of dimensions (parameters) \citep{winter_stellar_2020}.

We calculate this Mahalanobis phase space metric in two different ways in this work. First we calculate the Mahalanobis phase space metric for all stars with respect to all other stars and then we calculate the relative Mahalanobis phase space metric of host stars using the neighbourhood method from \citet{winter_stellar_2020}.

The neighbourhood method proceeds as follows. For each host star in turn we define a large neighbourhood (80 pc) and a local neighbourhood (40 pc). We check that the host star has at least 400 other stars within 40 pc of it; if there is not the host star is excluded from the analysis. If there are more than 600 stars within 40 pc we randomly pick 600 stars within the local neighbourhood of the host star. Then we calculate the Mahalanobis phase space metric for the host star with respect to all other stars within the local neighbourhood (not just the 400/600 found in the previous step). Then for each of the chosen 400/600 stars in the host stars local neighbourhood we calculate their Mahalanobis phase space metrics with respect to all other stars within 40 pc of them. Once the Mahalanobis phase space metric is calculated for all the chosen stars around the host star the results are median normalised to enable direct comparisons between different neighbourhoods.

\subsubsection{Mahalanobis phase space regimes}
\label{methods:regimes}
For each type of  neighbourhood (i.e. population of stars as defined above)  we split them into a low and high phase space regime following \citet{winter_stellar_2020}. 

To calculate if a exoplanet host star is in a low, ambiguous or high Mahalanobis phase space regime we fit two log-normals using Gaussian mixture modelling, as done by \citet{winter_stellar_2020}. We then calculate the probability, $P_{\rm {high}}$, that the host belongs to the high phase space regime. We use the same thresholds as \citet{winter_stellar_2020} where a $P_{\rm{high}} < 0.16$ corresponds to a low phase space regime, $P_{\rm{high}} > 0.86$ corresponds to a high phase space regime and anything between these values is said to belong to an ambiguous phase space regime. The threshold values are chosen to be one standard deviation apart.

Unlike the \citet{winter_stellar_2020} work we do not test if the simulations can be described using a single log-normal, due to the fact that our regimes are being defined for single star-forming regions. We have tried creating a Galactic disc population model by combining the outputs of different simulations. However, the results were not significantly different to a single stellar population from each individual $N$-body simulation.

\begin{figure*}
 \subfigure[30 au host, low phase space regime]{\includegraphics[width=0.45\linewidth,trim=2 2 2 2,clip]{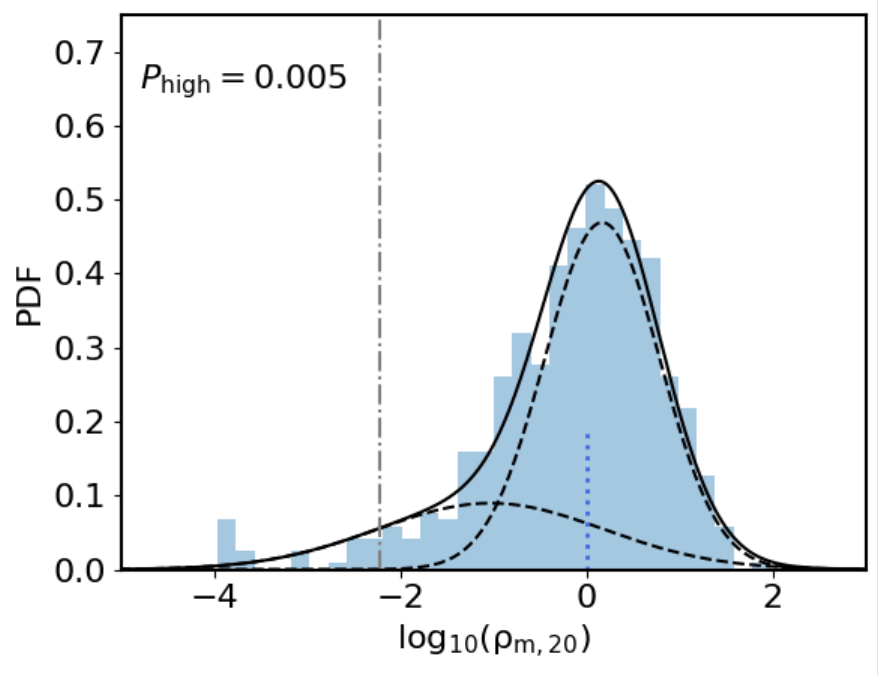}}
\hspace{-2.9cm}
 \subfigure[30 au host, high phase space regime]{\includegraphics[width=0.45\linewidth,trim=2 2 2 2,clip]{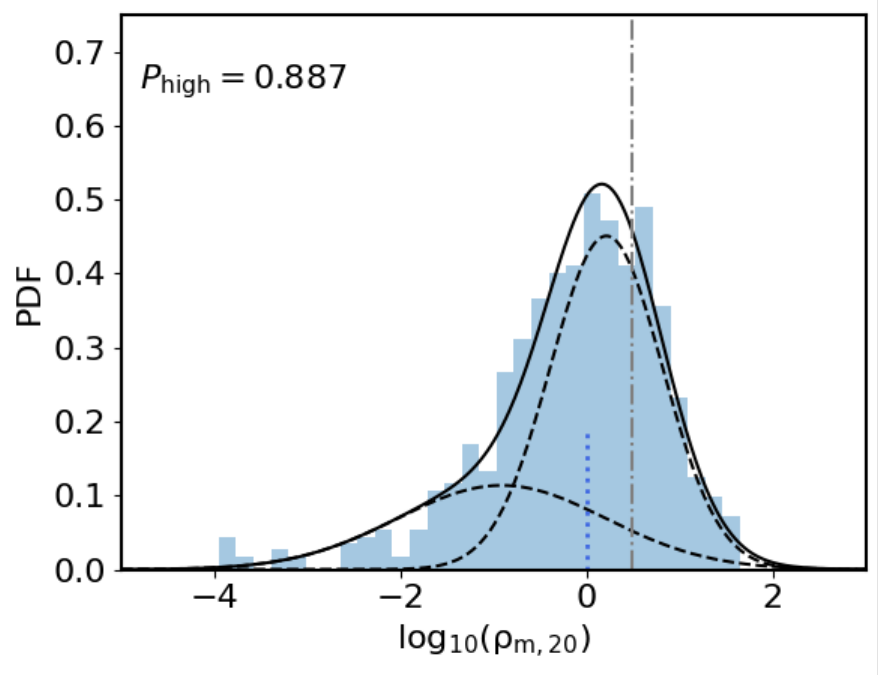}}
 \subfigure[5 au host, low phase space regime]{\includegraphics[width=0.45\linewidth,trim=2 2 2 2,clip]{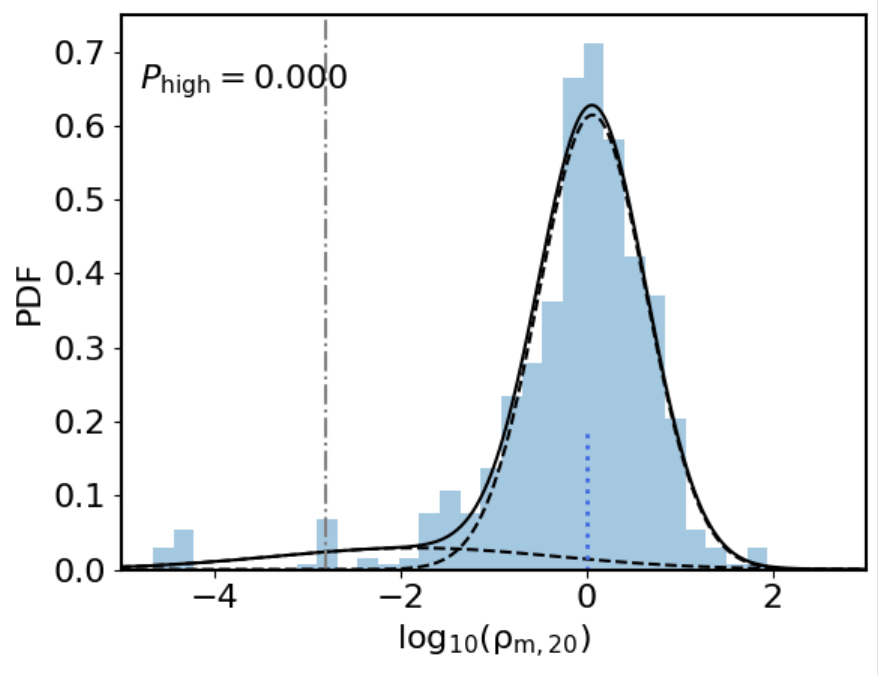}}
 \hspace{1.3cm}
 \subfigure[5 au host, high phase space regime]{\includegraphics[width=0.45\linewidth,trim=2 2 2 2,clip]{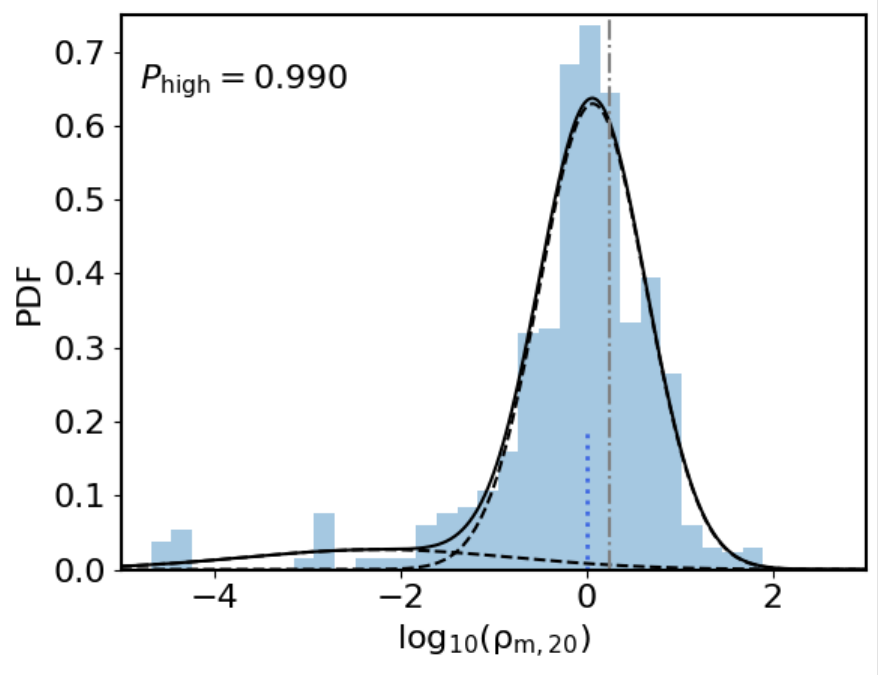}}
    
    \caption{Mahalanobis phase space metric distributions around host stars in different simulations. The grey dash--dotted line shows the Mahalanobis phase space metric for the host star. The black solid line shows the Gaussian mixture model consisting of two log-normals, with one corresponding to a low phase space regime and the other a high phase space regime. The dashed black lines show the component Gaussians and the blue dotted line which is always centred on zero highlights the median of the distribution after normalisation.}
    \label{fig:mahalanobis density regimes examples}
\end{figure*}

We show four examples of single host stars taken from simulations with different initial conditions in Figure~\ref{fig:mahalanobis density regimes examples}. Fig.~\ref{fig:mahalanobis density regimes examples} shows the distribution of Mahalanobis phase space metrics calculated for host stars and 600 randomly picked stars within 40 pc. These plots show examples of host stars in low (lefthand column) and high (righthand column) phase space regimes for simulations with planets at 30 au (top row) and 5 au (bottom row). The grey dash--dotted line shows the Mahalanobis phase space metric for the host star. The black solid line shows the Gaussian mixture model consisting of two log-normals, with one corresponding to a low phase space regime and the other a high phase space regime.

\section{Results}
\label{results}
In this section we present the Mahalanobis phase space evolution of \textit{N}-body simulations of 1000 stars with $\sim500$ planets with initial semi-major axes of 5 au or 30 au. We then show the number of times host stars are found in either the low, ambiguous or high Mahalanobis phase space regimes at $t = 0\,$Myr and $t = 10\,$Myr. We then show the number of perturbed host stars in each of the phase space regimes at $t = 10\,$Myr. 

\subsection{Definitions}
In this work we define planet `host' stars as being stars that are mutual nearest neighbours with a planet and that the total binding energy of the star--planet system is negative, meaning they are gravitationally bound. 
We define `former hosts' as stars that originally hosted a planet, but no longer do so in the current snapshot. We define `perturbed' hosts as stars whose planet's semi-major axis has changed by more than 10 percent of its initial value.

\subsection{Mahalanobis phase space evolution}
\begin{figure*}
 \subfigure[$a_i = 30\,$AU, high density]{\includegraphics[width=0.49\linewidth,trim=2 2 2 2,clip]{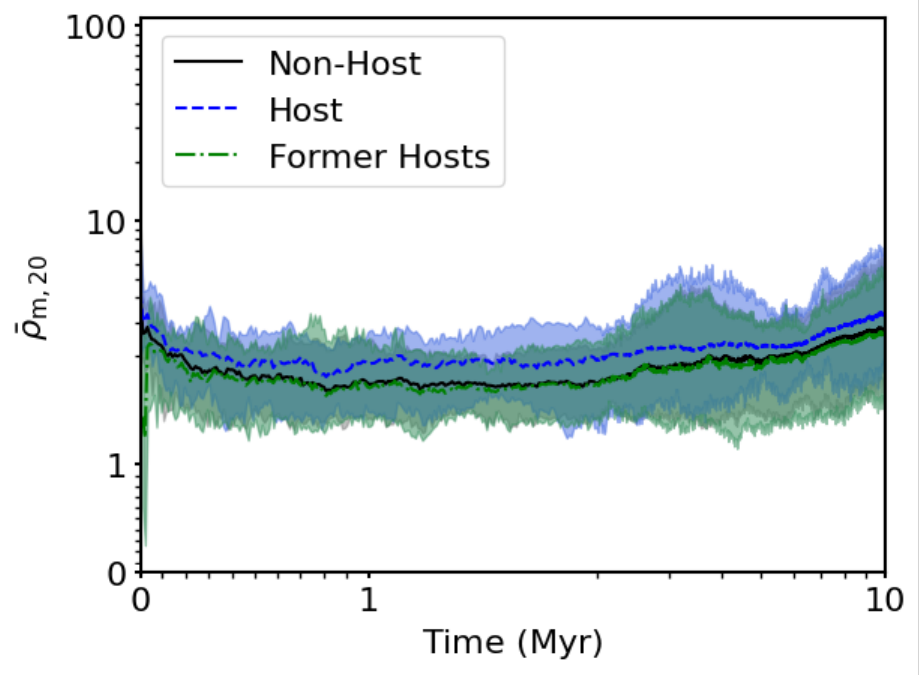}}
 \hspace{0.8pt}
 \subfigure[$a_i = 30\,$AU, low density]{\includegraphics[width=0.49\linewidth,trim=2 2 2 2,clip]{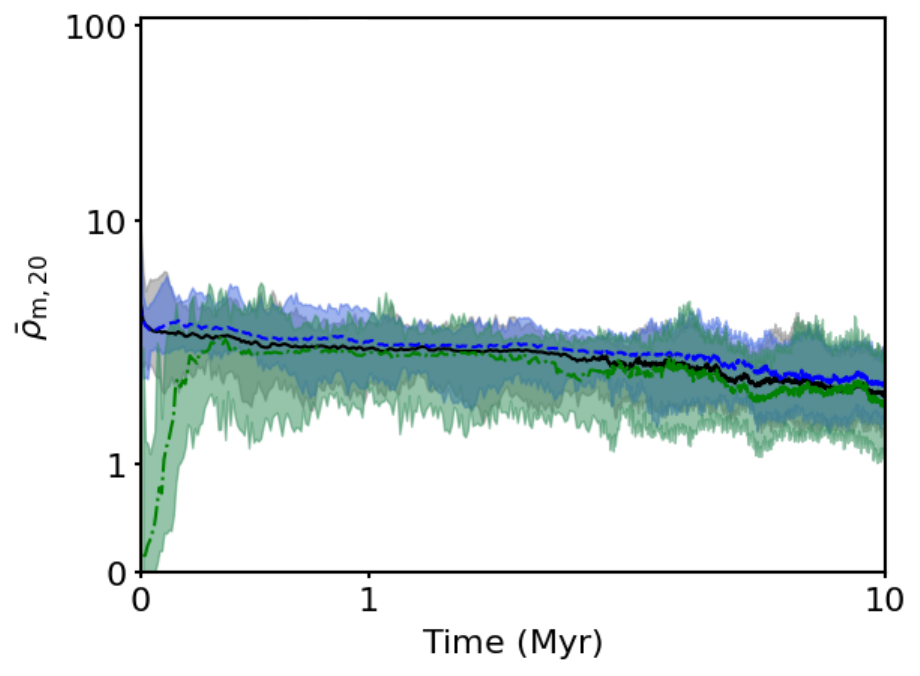}}
 \hspace{0.8pt}
 \subfigure[$a_i = 5\,$AU, high density]{\includegraphics[width=0.49\linewidth,trim=2 2 2 2,clip]{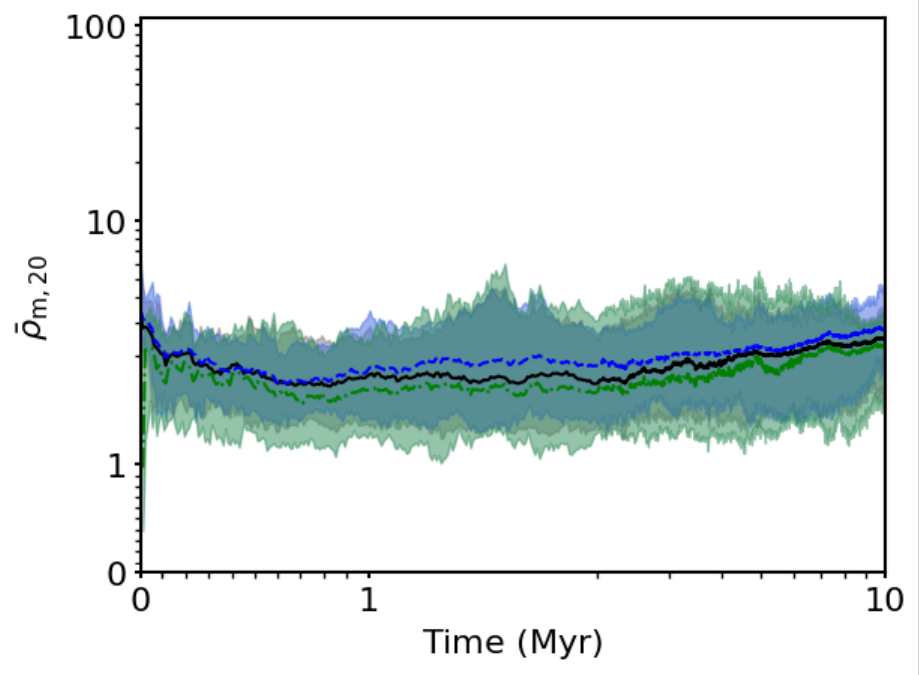}}
 \hspace{0.8pt}
 \subfigure[$a_i = 5\,$AU, low density]{\includegraphics[width=0.49\linewidth,trim=2 2 2 2,clip]{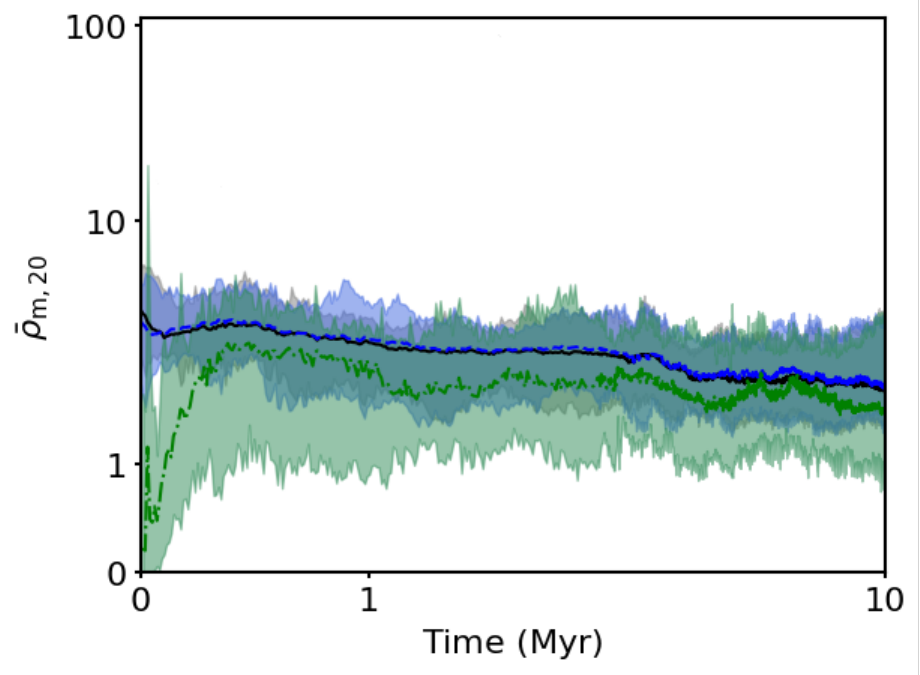}}
    
    \caption{The evolution of the Mahalanobis phase space metric against time for the four sets of simulations. The shaded  grey, blue and green areas show the range of mean Mahalanobis phase space metrics calculated across 20 simulations for non-host stars, host stars and former host stars, and the solid black, blue dashed and green dot-dashed lines show the median values for the 20 simulations. Simulations with planets with initial semi-major axes of 30\,au are shown in the top row and simulations with planets with initial semi-major axes of 5\,au are shown on the bottom row. The left hand column shows the high initial stellar density ($\tilde{\rho} \sim 10^4$\,M$_{\odot}\,$pc$^{-3}$) simulations and the right hand column shows the low initial stellar density ($\tilde{\rho} \sim 10^2$\,M$_{\odot}\,$pc$^{-3}$) simulations.}
    \label{fig:mahalanobis density over time}
\end{figure*}

In Figure~\ref{fig:mahalanobis density over time} we show the evolution of the mean 6D (position and velocity information) Mahalanobis phase space metric (as defined by Eqn.~\ref{mahalanobis_eqn}) of non-host stars (stars with no gravitationally bound planet), host stars (stars with at least one gravitationally bound planet) and former host stars (stars that have lost their planet). Fig.~\ref{fig:mahalanobis density over time}(a) shows the evolution of the mean Mahalanobis phase space metric for high initial density ($\tilde{\rho} \sim 10^4$\,M$_{\odot}\,$pc$^{-3}$) simulations with planets initially on 30 au orbits. The host stars (blue dashed line) have a higher Mahalanobis phase space metric, whereas the non-hosts and former hosts have a slightly lower phase space metric. However, the uncertainties between different realisations of the same initial conditions overlap to such an extent that observing one snapshot in time would make differentiating between host, non-host and former host stars, difficult.  

In Fig.~\ref{fig:mahalanobis density over time}(b) we show the evolution of the Mahalanobis phase space metric for the initially low density ($\tilde{\rho} \sim 10^2$\,M$_{\odot}\,$pc$^{-3}$), but where the planets are again on 30\,au orbits. In these simulations the former host stars initially have much lower mean Mahalanobis phase space metrics. We attribute this to the effects of low number statistics as early on in the simulations there are no former hosts, because interactions that disrupt planetary systems take longer to occur in these low density simulations. 

We now repeat the above analyses but for simulations where the planets are initially all on 5\,au orbits. Fig.~\ref{fig:mahalanobis density over time}(c) shows the evolution of the Mahalanobis phase space metric for the high density simulations. There are slight differences in the evolution of the metric depending on whether we are looking at planet hosts, non-hosts or former hosts, with the former hosts having the lowest phase space metric values. However,  there is significant overlap in the uncertainties in the metrics between the host, non-host, and former host stars. 

In Fig.~\ref{fig:mahalanobis density over time}(d) we show the evolution of the mean Mahalanobis phase space metric for the low density simulations. There is no significant difference between the evolution of the Mahalanobis phase space metric  for non-host and host stars, but there is a difference in the evolution of the metric for the former hosts, in that the former host stars consistently have lower minimum mean Mahalanobis phase space metrics.

We interpret the lower Mahalanobis phase space metric for the former host stars as being due to is due to the kinematics of the former host stars. 

In the low density regions stars will undergo fewer dynamical interactions, thus the velocities of the majority of stars (including the planet hosts) are unlikely to be significantly altered. However, planets initially on (say) 5\,au orbits will be strongly bound to their stars, and therefore any interaction that is energetic enough to remove that planet will significantly alter the velocity of that star. We believe this is the reason for the former hosts displaying slightly lower Mahalanobis phase space metrics in the low density simulations (panels b and d of Fig.~\ref{fig:mahalanobis density over time}), and the high density simulations where the planets are initially tightly bound (Fig.~\ref{fig:mahalanobis density over time}c).

In the high density simulations where the planets are on more weakly bound orbits (Fig.~\ref{fig:mahalanobis density over time}a), more stars undergo energetic interactions, and these interactions are more likely to disrupt planets on 30\,au orbits. For this reason, the Mahalanobis phase space metric is not significantly different for the former host stars.

The difference in the kinematics for planets in different density regions is further highlighted by the mean number of former hosts in each set of  simulations. For the simulation sets with planets at 30\,au we find that by the end of the simulations ($t = 10\,$Myr) there are 169 and 308 former hosts for low and high stellar density regions, respectively. We see a similar trend for the simulation sets with planets initially at 5\,au, with 76 and 163 former hosts at the end of the simulations for low and high stellar densities, respectively. The lower number of former hosts for planets initially at 5\,au in both low and high stellar densities is due to planets at 5\,au being more strongly gravitationally bound to their host stars, making them harder to remove via external interactions with other stars \citep{heggie_binary_1975,1975AJ.....80..809H,2006ApJ...640.1086F,2013MNRAS.432.2378P}.

This result is only discernible because we know everything about our simulations, i.e.\,\,we know exactly which stars have lost planets in dynamical interactions. This is not possible observationally; however a statistical method could be developed to  estimate the number of stars likely to have lost planets, but this method would require assumptions about the typical planetary system architecture and the initial density of the star-forming region.

\subsection{Host star Mahalanobis phase space regimes}

We now examine the Mahalanobis phase space metric distribution in more detail at $t = 0$\,Myr (before any dynamical evolution has taken place) and $t = 10$\,Myr (the end of our simulations). Based on their Mahalanobis phase space metric, we divide the stars into three regimes (as detailed in Section~\ref{methods:regimes}); low, ambiguous and high (phase space metric).

\begin{figure*}
 \subfigure[Host star phase space regimes at $t = 0\,$Myr]{\includegraphics[width=0.49\linewidth,trim=2 2 2 2,clip]{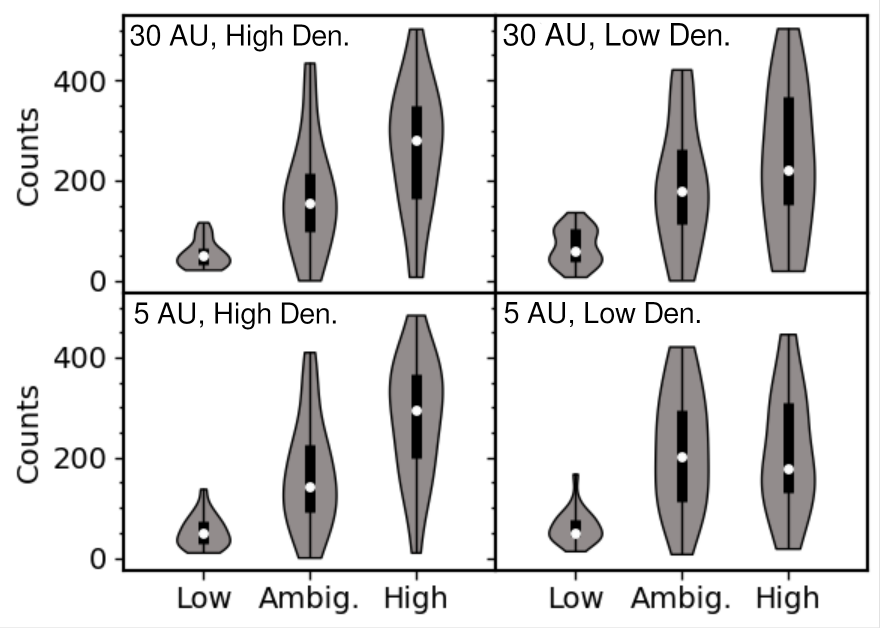}}
 \hspace{0.8pt}
 \subfigure[Host star phase space regimes at $t = 10\,$Myr]{\includegraphics[width=0.49\linewidth,trim=2 2 2 2,clip]{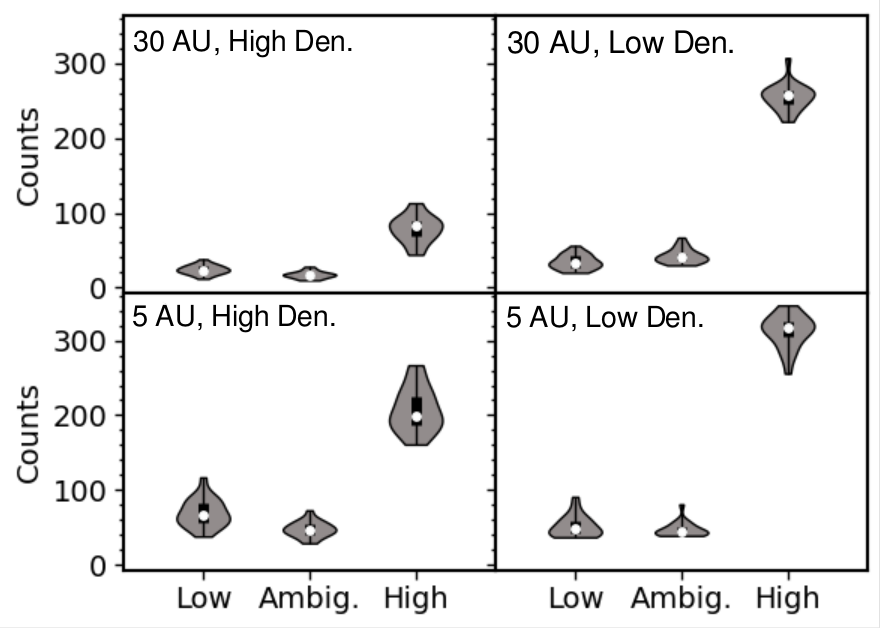}}
    \caption{Violin plots of the number of host stars in either low, ambiguous or high Mahalanobis phase space regimes at 0 Myr (panel a) or 10 Myr (panel b). These plots show the counts across all 20 simulations in the four different sets we use. The white dot is the median count, the black bar is the interquartile range, the thin black line shows the full range of values and the thickness of the plot corresponds to the probability of finding a particular value. The top left plot shows the counts for hosts with planets initially at 30\,au in high stellar density regions ($\tilde{\rho} \sim 10^4$\,M$_{\odot}\,$pc$^{-3}$), and the top right plot shows the same but for low stellar density regions ($\tilde{\rho} \sim 10^2$\,M$_{\odot}\,$pc$^{-3}$). The bottom left plot shows the counts for hosts with planets initially at 5\,au in high stellar density regions, and the bottom right plot shows the same but for low stellar density regions.}
    \label{fig:initial final snap density regimes}
\end{figure*}

We show the initial and final Mahalanobis phase space metric distributions in Fig.~\ref{fig:initial final snap density regimes} of host stars across the 20 simulations for each set of stars that are in either low, ambiguous or high Mahalanobis phase space regimes. 

We find that initially most host stars are either in the ambiguous or high phase space regime (Fig.~\ref{fig:initial final snap density regimes}a), with a similar number of hosts in both of these regimes.  This is expected as these simulated star-forming regions are initially highly spatially and kinematically substructured. This means that stars close to each other in physical space have small velocity dispersions, which leads to a high Mahalanobis phase space metric. 

We show the final ($t = 10$\,Myr) distributions in Fig.~\ref{fig:initial final snap density regimes}(b).  Most host stars across all four sets of simulations are in the high Mahalanobis phase space regime. 


We interpret the far higher numbers of host stars in the high Mahalanobis phase space regime at 10\,Myr as being due to stars that remain hosts not having undergone as much dynamical processing as former hosts. If they had undergone dynamical encounters, they  would likely have lost their planet, becoming former hosts. Conversely, the general  population of stars will have velocities that have been changed due to dynamical interactions which will manifest as an overall decrease in the Mahalanobis phase space metric. This results in more host stars residing in the high Mahalanobis phase space  regimes.

\subsubsection{Host stars with perturbed planets}

\begin{figure*}
\centering
 \subfigure[Perturbed host star phase space regimes $t = 10\,$Myr]{\includegraphics[width=0.49\linewidth,trim=2 2 2 2,clip]{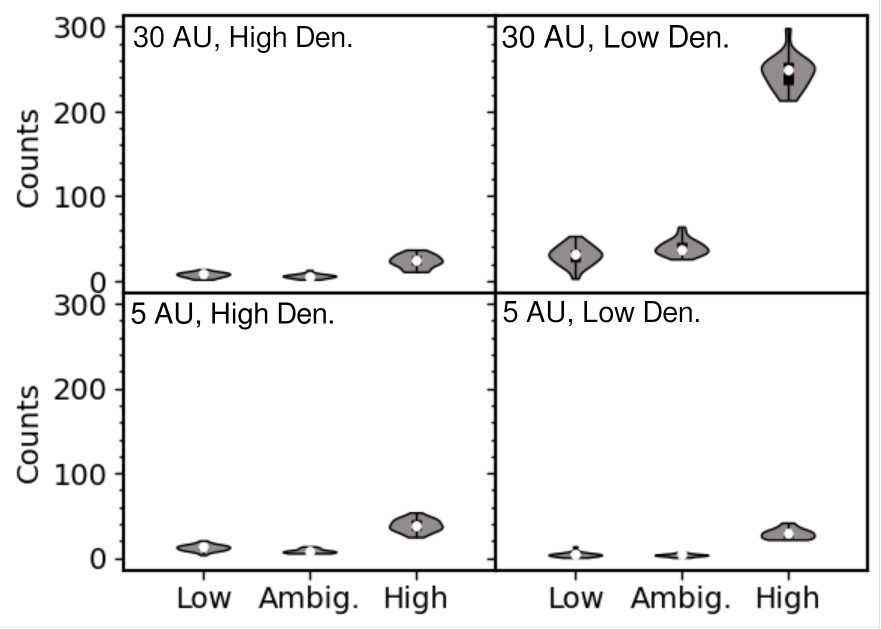}}

    \caption{Violin plots of the number of perturbed host stars (stars whose planets' semi-major axes have changed by $\pm$10 per cent) after 10\,Myr of dynamical evolution in either low, ambiguous or high Mahalanobis phase space regimes across all 20 simulations. In all panels, the white dot is the median count, the black bar is the interquartile range, the thin black line shows the full range of values and the thickness of the plot corresponds to the probability of finding a particular value. The top-left plot shows the counts for hosts with planets initially at 30\,au in high stellar density ($\tilde{\rho} \sim 10^4$\,M$_{\odot}\,$pc$^{-3}$) regions, top-right shows the same but for low stellar density ($\tilde{\rho} \sim 10^2$\,M$_{\odot}\,$pc$^{-3}$)  regions. The bottom left plot  shows the counts for hosts with planets initially at 5\,au in high stellar density regions, and bottom right shows the same but for low stellar density regions.}
    \label{fig:density regime pert hosts}
\end{figure*}

We now plot the Mahalanobis phase space metric distributions for host stars whose planets have been perturbed (i.e.\,\,the semimajor axis has changed by $\pm$10\,per cent of its initial value, after 10\,Myr of dynamical evolution. Fig.~\ref{fig:density regime pert hosts} shows the number of perturbed hosts in either the low, ambiguous or high Mahalanobis phase space regime. 

In all sets of simulations, the trends are similar to the distributions for all planet host stars we present in Fig.~\ref{fig:initial final snap density regimes}(b). This is to be expected, as the perturbed hosts represent a subset of all of the host stars. 


However, we see far more perturbed hosts in the high Mahalanobis phase space regimes  in low density ($\tilde{\rho} \sim 10^2$\,M$_{\odot}\,$pc$^{-3}$) simulations where the planets have initial semi-major axes of 30\,au, than in any other simulations. There are fewer perturbed hosts in the high density simulations due to the interactions in the simulations being more likely to produce free floating planets, whereas the low density simulations are less likely to remove planets and instead just perturb them (this can be seen in the difference between distributions in the top left and top right of Fig.~\ref{fig:density regime pert hosts}). 

Conversely, there is very little difference in the Mahalanobis phase space distributions when the host stars' planets are at 5\,au; this is due to very few planets being perturbed at low stellar densities.



\section{Discussion}
\label{discuss}

Our main result is that the Mahalanobis phase space `density' -- a metric that incorporates information on the position and velocity of each star and provides a dimensionless distance from the average in a distribution -- of planet host stars only weakly depends on the dynamical history of the star--planet system. When we plot the evolution of the Mahalanobis phase space metric over time, stars whose planets have been liberated by dynamical interactions have a slightly lower Mahalanobis phase space metric, but one that is within the spread of the metric for stars who still retain their planets, and stars that have never hosted planets.

When we analyse the Mahalanobis phase space metric for `neighbourhoods' of stars (similar to the approach in \citet{winter_stellar_2020}) we see a clear preference for planet host stars to reside in areas where the phase space metric (`density') is high, regardless of the initial \emph{physical} density of the star-forming region. Rather than the physical density, the high phase space metric is actually dominated by the kinematics of the star-forming region. 

Star-forming regions form with spatial and kinematic substructure \citep{larson_turbulence_1981,gomez_spatial_1993,cartwright_statistical_2004,sanchez_spatial_2009,alfaro_looking_2016}, and our simulations are designed to mimic this before any dynamical evolution takes place. For this reason, more stars reside in a high or ambiguous Mahalanobis phase space before dynamical evolution. Once dynamical evolution has occurred, the majority of stars that retain planets reside in high Mahalanobis phase space (Fig.~\ref{fig:initial final snap density regimes}). This is because stars that do not undergo interactions that would significantly alter the local kinematic substructure are likely to retain their planets. 

\citet{winter_stellar_2020} interpret the high Mahalanobis phase space densities they find around Hot Jupiter host stars as evidence that these objects formed in a high physical density environment, which has caused the formation of the Hot Jupiters from dynamical processes, such as a direct encounter with a passing stars \citep{parker_quanz2012,2024arXiv240111613B}, planet--planet scattering \citep{2009ApJ...699L..88R} or from von~Zeipel-Lidov-Kozai cycles \citep{2003ApJ...589..605W,malmberg2007,2009MNRAS.397.1041P}.

However, we find that planet host stars that do form in dense stellar environments will have a high Mahalanobis phase space metric simply because the kinematic signsture from their birth environment is retained. To this end, we expect planet host stars in all regions where there is some degree of kinematic coherence \citep[e.g. young moving groups,][]{2004ARA&A..42..685Z,2015ApJ...798...73G,2019AJ....157..234S} to exhibit high a Mahalanobis phase space metric. 

Planets that are perturbed by dynamical encounters with passing stars, but remain in orbit around their parent star, will attain higher than average Mahalanobis phase space metric values, but not significantly more so than planets that have not been perturbed. Again, this is due to the Mahalanobis phase space metric being dominated by the velocities of the stars, rather than the spatial density of their birth environment.   

We therefore cannot determine the dynamical history of a Hot Jupiter host star using its local Mahalanobis phase space metric, and the question of whether Hot Jupiters form due to disc migration processes, or external dynamical encounters remains open.

Our results appear complementary to those those of \citet{adibekyan_stellar_2021} and \citet{mustill_hot_2022}, who posit that a high Mahalanobis phase space metric is indicative of differences in the kinematics of Hot Jupiter host stars compared to stars without Hot Jupiters, something these authors attribute to age biases. 

Finally, we emphasise that our work is subject to several important caveats and assumptions. First, unlike \citet{winter_stellar_2020}, who analyse the Galactic field population (which is presumably comprised of many different star-forming regions), our analysis is on individual star-forming regions. In a set of exploratory simulations, we find that our results are similar if we create a composite dataset of different snapshots from e.g. different simulations at different ages, and perform a similar analysis on these stellar populations. 

Second, our simulations do not include multi-planet systems, so we cannot quantify whether dynamical instabilities on the outer planet would induce further instabilities on the inner planets \citep[e.g.][]{2019MNRAS.489.2280F}, potentially leading to more Hot Jupiters. 

Finally, we do not include the effects of dissolution of our star-forming regions due to feedback from massive stars \citep{goodwin_stellar_2006,shukirgaliyev_impact_2017}. If this process is important, it is likely to ``freeze in" the velocities of the planet host stars, and reduce the number of perturbed planets, thus not affecting our interpretation. 


\section{Conclusions}
\label{conclude}

We have calculated the 6D Mahalanobis phase space `density' metric for sets of $N$-body simulations with different initial physical densities (low density $\sim 10^{2}\,$M$_{\odot}\,$pc$^{-3}$ and high density $\sim 10^{4}\,$M$_{\odot}\,$pc$^{-3}$) and different semi-major axes of Jupiter mass planets (all planets at 5\,au or 30\,au). We then compare the Mahalanobis phase space metric for stars that have lost their planets, stars that have retained their planets, and stars that have retained their planets but whose semi-major axis has changed by more than $\pm$10\,per cent. Our conclusions are as follows.
\begin{enumerate}
    \item The 6D Mahalanobis phase space evolution for different subsets of stars (planet hosts, non-hosts and former hosts) is similar for simulations with different initial physical densities. 

    \item We find that host stars are predominantly found in high Mahalanobis phase space regimes at the end of all the simulations ($t = 10\,$Myr), regardless of the initial physical density. This is due to the host stars retaining some of the kinematic substructure from their formation environment.

    \item We find that host star whose planets' semi-major axes have changed by more than 10 per cent (perturbed host stars) are also found in high Mahalanobis phase space regimes regardless of the initial physical density. 

\end{enumerate}

Based on our results, we would not expect dynamical interactions in a planet host star's birth environment to be responsible for the high Mahalanobis phase space densities, as hypothesised in \citet{winter_stellar_2020}. Conversely, we find that unperturbed planet host stars are most likely to be found in high Mahalanobis phase space, because they retain some kinematic substructure from their birth star-forming region, which tends to be erased for stars that experience strong or repeated dynamical encounters. 


\begin{acknowledgments}
GABS acknowledges a University of Sheffield publication scholarship. RJP acknowledges support from the Royal
Society in the form of a Dorothy Hodgkin Fellowship. ECDP acknowledges funding from UKRI in the form of an STFC PhD scholarship. For the purpose of open access, the author has applied a Creative Commons Attribution (CC BY) licence to any Author Accepted Manuscript version arising.
\end{acknowledgments}



\software{Matplotlib 3.3.4 \citep{hunter_matplotlib_2007},  
          Numpy \citep{harris_array_2020}, 
          SciPy \citep{virtanen_scipy_2020}
          }





\bibliography{bibliography_exohost_stars}
\bibliographystyle{aasjournal}



\end{document}